\documentstyle[twocolumn,aps,prl]{revtex}
\def\Dsb#1{{\langle}#1}       % standard bra
\def\Dsk#1{#1{\rangle}}       % standard ket
\begin{document}
\hbadness=1500
\title{Floydian trajectories for stationary systems: a modification for bound states}
\draft
\author{M. R. Brown}
\address{Theoretical Physics Research Unit, Birkbeck College, University of London,
Malet Street, London WC1E 7HX, England}

\date{\today}
\maketitle
\begin{abstract}
The Floydian trajectory method of quantum mechanics and the
appearance of microstates of the Schr\"{o}dinger equation are
reviewed and contrasted with the Bohm interpretation of quantum
mechanics. The kinematic equation of Floydian trajectories is
analysed in detail and a new definition of the variational
derivative of kinetic energy with respect to total energy is
proposed for which Floydian trajectories have an explicit time
dependence with a frequency equal to the beat frequency between
adjacent pairs of energy eigenstates in the case of bound systems.
In the case of unbound systems, Floydian and Bohmian trajectories
are found to be related by a local transformation of time which is
determined by the quantum potential.
\end{abstract}

% body of paper here

\section{Introduction}
\label{sec:introbofl}

In a series of papers \cite[and references therein]{Floyd00a},
Floyd has developed a trajectory representation of quantum
mechanics which, while related to the Bohm interpretation, is
distinct in both in its trajectories and its interpretation. Whilst
the Floydian method has been strongly advocated and well
exemplified by its proponent, in comparison with the Bohm
interpretation it seems not to have received as wide critical
attention in the literature as perhaps it deserves. Here we attempt
to partly redress the balance by briefly reviewing the Floydian
method and then critically examining the basis of the kinematic
equation of its trajectories. We refer to the Bohm interpretation
to stimulate the critical analysis.  Floyd has made a detailed
comparison of the two approaches in~\cite{Floyd00}.

Two novel aspects of the Floydian method are: 1) the identification
of microstates of the Schr\"{o}dinger equation and 2) a kinematic
equation for trajectories which is derived from the Hamilton-Jacobi
method of classical mechanics.  The method is based upon the direct
solution of the Quantum Stationary Hamilton-Jacobi Equation (QSHJE)
rather than upon the solution of the corresponding Schr\"{o}dinger
equation and has, thus far, only been developed for 1-dimensional
and pseudo-2-dimensional stationary systems.  In particular, it has
been applied to the finite potential step, the semi-infinite step
(in 2-dimensions)~\cite{Floyd00a}, the infinite~\cite{Floyd99} and
the finite square potential well~\cite{Floyd99a}. In early work,
solutions of the QSHJE were obtained numerically~\cite{Floyd82} but
following the discovery of the form of a general solution in
1-dimension~\cite{Floyd86}, attention has been directed towards
studying analytical solutions. The results can be extended to
higher dimensions if separation of variables is permitted. However,
closed form solutions are not known in the general case and whilst
it has been briefly discussed in principle~\cite{Floyd96}, the
general case of higher dimensions remains untried in practice.

More recently, the work of Floyd has been set in a wider context by
the work of Faraggi and Matone~\cite{Matone00c} who derive and
generalise (to higher dimensions) the QSHJE on the basis of a
postulated equivalence principle: that all physical systems are
equivalent under coordinate transformations.  This connection is of
interest, not only because it provides a wider context for the work
of Floyd but also because it motivates a connection between the
quantum potential and coordinate transformations or deformations of
the geometry of space.

Whilst we briefly review the Floydian method and the basis of the
appearance of microstates of the Schr\"{o}dinger equation, our
primary focus is upon its trajectory representation. We critically
assess the application of the Floydian trajectory method to bound
systems.  By working in a 3-dimensional coordinate representation
we also highlight necessary further assumptions for its application
in higher dimensions.  This is in contrast to the principally
1-dimensional approach of Floyd.

A point of particular attention, is the appearance of the energy
derivative of the kinetic energy in the kinematic equation of
Floydian trajectories, especially in quantum systems with discrete
energy spectra.  Such a derivative is well defined (in fact it is
unity) for classical conservative systems, since the transition
between one 'stationary' energy state of a given system to another
is continuous.  However, in bound quantum systems, not only is the
energy spectrum discrete (complicating the definition of an energy
derivative), but also any attempt to make a transition between the
states in order to define a continuous derivative with respect to
energy whilst at the same time continuing to satisfy the
Schr\"{o}dinger equation and its boundary conditions faces two
problems: 1) states between the eigenstates are time dependent so
resulting in a time and space dependent energy derivative 2) there
is a multiplicity of such derivatives from which to choose. We
explore these problems, (thus far investigated by Floyd only
numerically with respect to the harmonic
oscillator~\cite{Floyd82x}), for stationary systems in general and
examine their relevance to the energy variation of the QSHJE as
used by Floyd. These issues do not arise in systems with continuous
energy spectra where the Floydian method has been used to
investigate barrier reflection and tunnelling problems.

The following sections proceed as follows. By way of establishing a
comparative base, we first briefly review the equations of the Bohm
trajectory method for the general case and subsequently for
stationary systems.  After briefly reviewing the work of Faraggi
and Matone, we then review the Floydian method and the appearance
of microstates contrasting the approach with that of Bohm.
Deferring until later the kinematic equation of Floydian
trajectories, we then derive, for its use, the variational
derivative of the kinetic energy with respect to the total energy
for stationary systems revealing its space-time dependence and its
ambiguity. Taking a more general approach than Floyd, we then
derive the kinematic equation of a Floydian trajectory from the
variational derivative of the action with respect to the total
energy, described by Floyd as the epoch $\tau$.  We then use our
results to demonstrate that Bohmian and Floydian trajectories are
related by different assumptions about $\tau$ and we are led to the
idea of time deformation by the quantum potential. After briefly
comparing applications to different types of potential, we finally
draw together our results in the conclusions.

\section{The Bohm trajectory method}
\label{Bohtm}

In the Bohm approach~\cite{BH93} the quantum Hamilton-Jacobi
equation is derived, along with the equation for conservation of
probability, from the Schr\"{o}dinger equation and its complex
conjugate by using the polar form of the wave function $\psi
=R(q,t) \exp(i
\frac{S(q,t)}{\hbar})$. For stationary systems, $R = R(q)$ and
$S(q,t)=W(q)- Et$, where $W(q)$ is Hamilton's characteristic
function. In the Bohmian method, one obtains $S(q,t)$ directly from
the wave function solution of the Schr\"{o}dinger equation. The
trajectory kinematics are derived from a classical form of the
probability current density identified in the probability
conservation equation.  This method is therefore rooted in the
probabilistic interpretation of the wave function.

In the coordinate representation, the substitution $\psi
=R(q,t) \exp(i \frac{S(q,t)}{\hbar})$ in the Schr\"{o}dinger equation yields:
\begin{equation}
\frac{1}{2m} {\left( \nabla{S} \right)}^{2} +
V(q) + Q(q) = -\frac{\partial S}{\partial t}
\label{eq:Boht1}
\end{equation}
and
\begin{equation}
\frac{\partial {R}^{2}}{\partial t} + \nabla \cdot
\left({\frac{R^{2}}{m} \nabla S }\right) = 0
\label{eq:Boht2}
\end{equation}
where
\begin{displaymath}
Q = - \frac{{\hbar}^{2}}{2mR} {\nabla}^{2} R
\end{displaymath}
is the quantum potential.  From the probability conservation
equation~(\ref{eq:Boht2}), one identifies the probability current
density $j = {\frac{R^{2}}{m} \nabla S}$. Using the classical form
$j = \rho u$ (where $\rho$ is the density and $u$ the velocity of a
flow) one then identifies $\dot{q}
= \frac{1}{m} \nabla S$ as the streamline velocity of the
probability density current which upon integration yields a
trajectory.  If the density of such trajectories is at any time
$R^{2}$, then equation~(\ref{eq:Boht2}) ensures that it is always
so.  We also observe from equation~(\ref{eq:Boht1}), that with the
momentum defined by $p = \nabla S$, this equation is the
Hamilton-Jacobi equation but for the quantum potential $Q(q)$. One
may therefore consistently assign $p = m\dot{q}$.  The latter
assignment is strictly true only for conservative systems in a
cartesian coordinate representation.

In the case of stationary systems, $S(q,t)=W(q)- Et$ and so the
above equations reduce to
\begin{equation}
\frac{1}{2m} {\left(\nabla W \right)}^{2} +
V(q) + Q(q) = E
\label{eq:Boht3}
\end{equation}
and
\begin{equation}
\nabla \cdot
\left({\frac{R^{2}}{m} \nabla W }\right) = 0
\label{eq:Boht4}
\end{equation}
with the trajectory equation
\begin{equation}
\dot{q} = \frac{1}{m} \nabla W.
\label{eq:Boht5}
\end{equation}

From the above, we see that for a stationary system of a given
energy E, a Bohm trajectory is defined by a square integrable
solution of the Schr\"{o}dinger equation (to secure the concept of
probability) and an initial condition $x_0$.  Moreover, owing to
the single-valuedness of the wave function, Bohm trajectories for a
given eigenstate do not cross.  For stationary systems with real
wave functions, Bohm trajectories are stationary; a point of
concern to Einstein who observed that, in the classical $\hbar
\rightarrow 0$ limit, the classical motion is not obtained ~\cite{Holl93},
p243.  Floyd~\cite{Floyd94} has observed that the use of a
trigonometric (rather than complex polar) ansatz resolves this
artifact.  We do not pursue this further here.

\section{The quantum stationary Hamilton-Jacobi equation}
\label{eq:floyt}
%[17|D] [1|E] [2|E] [3|E] [13|J] q-p/9907092,q-p/0009070, q-p/9708026
% q-p/9707051

\subsection{The quantum stationary Hamilton-Jacobi
equation via an equivalence principle}
\label{sec:eqpri}
%[63|K]

More recently than Floyd's early work, Faraggi and
Matone~\cite{Matone99a,Matone99b,Matone99c,Matone98a,Matone98b,Matone00a,Matone00b,Matone00c}
have independently derived (and generalised to higher dimensions)
the QSHJE from a postulated equivalence principle: that all
physical systems are equivalent under coordinate transformations.
This work extends the implications of Floyd's work on the QSHJE to
a connection between quantum mechanics and
gravity~\cite{Matone00c}. In the following, we summarise those key
points of the work of Matone et al. that have a bearing on our work
here:

\begin{itemize}
  \item A key corollary of the equivalence postulate is that there
  must exist a trivialising coordinate transformation $q\to q_0$,
  defined by the relation between the respective Hamilton's
  characteristic functions (reduced actions) $W(q)=W_0(q_0)$, which
  reduces any system to that of the free particle~\cite{Matone99c}:
  \begin{displaymath}
  {\mathcal{W}} ( q ) \to {\mathcal{W}}_{0} ( q_0 )
  \end{displaymath}
  where ${\mathcal{W}} = V ( q ) - E$ and ${\mathcal {W}}_{0}
  \equiv 0$.

  \item The quantum potential term in the QSHJE is a direct
  consequence of the postulated universal existence of the
  trivialising coordinate transformation.  In particular, the
  absence of the quantum potential (as in classical mechanics)
  prohibits the existence of such a transformation to a frame in
  which a system is 'at rest'.  The existence of the quantum
  potential therefore enables the removal of this privileged frame.

  \item The quantum potential is expressed in terms of the
  Schwarzian derivative of the characteristic function or reduced
  action $W_{\mu}$ (see section~\ref{sec:qshjesol}) which must
  therefore never be a constant in order for the trivialising map
  to exist. To guarantee this, a general solution of the
  Schr\"{o}dinger equation should be expressed in the form $\psi =
   R_{\mu} (A \exp{(i W_{\mu} / \hbar)} + B \exp{(-i W_{\mu} /
  \hbar)})$ rather than in the Bohmian form $\psi = R \exp{(i W /
  \hbar)}$ for real $R$, $W$, $R_{\mu}$, and $W_{\mu}$ with $A$ and
  $B$ as complex constants.  In the case of real wave functions,
  this implies that $A$ and $B$ must be of equal magnitude and,
  modulo an integer factor $2\pi$, of opposite signed phase so that
  a real wave function is the sum of a Bohmian polar form and its
  complex conjugate.  (Here and hereafter, we distinguish the
  characteristic function, $W_{\mu}$, as a direct solution of the
  QSHJE from the the reduced action, W, appearing in the phase of
  the wave function, as used in the Bohm formulation.)

  \item A general principle is that a general solution of the
  Schr\"{o}dinger equation is always has two independent solutions.
  In general, they are not square integrable.  However, under the
  special conditions for eigenfunctions, one and only one of them
  may be a square-integrable function on the real line.  The
  Copenhagen interpretation selects the latter solution in order
  that the wave function be interpreted as a probability amplitude.

  \item In the case of the QSHJE, a condition for the existence of
  the Schwarzian derivative is that the total energy $E$ must take
  a value such that the QSHJE corresponds to a Schr\"{o}dinger
  equation having a wave function that is a square-integrable
  function on the real line ie $E$ is an eigenvalue of the
  corresponding Schr\"{o}dinger equation.   Thus, in solving the
  QSHJE, quantisation arises independently of the Copenhagen
  interpretation demand for square-integrable solutions.

  \item Time does not appear explicitly in the QSHJE and so the
  time parameterisation is introduced, after the solutions of the
  QSHJE are obtained, in the form of trajectories defined using the
  Floydian ansatz $t -\tau = \partial W_{\mu} / \partial E$.  In
  classical mechanics this is equivalent to the kinematic relation
  $p=m\dot{q}$ in which time parameterisation is implicit.
\end{itemize}

The work of Faraggi et al. draws upon the requirement for
satisfactory classical limits (expressed as $\hbar \rightarrow 0$)
and through its insistence upon a non-constant reduced action,
claims to have side-stepped Einstein's objection that the Bohm
interpretation does not have the correct classical limit (in the
aforementioned sense) for bound systems such as the quantum
harmonic oscillator. We also see that the quantum potential derived
from the equivalence principle and used in Floyd's trajectory
method does not necessarily equal that of the Bohm interpretation,
for the latter is based upon the polar decomposition of only square
integrable wave-functions whereas the former are based on the
direct solution of the QSHJE and correspond to more general
solutions of the Schr\"{o}dinger equation.

In the next section we focus upon the specific treatment of the
QSHJE by Floyd.

\subsection{Overview of the Floyd method}
\label{sec:ovfloyd}

The Floydian
approach~\cite{Floyd82,Floyd84,Floyd86,Floyd94,Floyd96,Floyd99a,Floyd00}
originated from an alternative ansatz to the Bohm polar form for
solutions of the Schr\"{o}dinger equation, which is nevertheless
consistent with the second Bohm (probability conservation)
equation~(\ref{eq:Boht4}). Substituted in the Schr\"{o}dinger
equations, the Floyd ansatz results in a single third order quantum
stationary Hamilton-Jacobi (QSHJE) equation for the quantum form of
Hamilton's characteristic function $W_{\mu}(q)$. For a given energy
eigenvalue of the Schr\"{o}dinger equation, $W_{\mu}(q)$ is not a
unique solution of the QSHJE and this leads to the idea that the
square integrable wave function for bound states is not exhaustive
but has associated microstates.  As seen in the previous section,
this conclusion has been more generally substantiated by the work
of Farragi and Matone.  Whilst the Floydian
ansatz~\cite{Carroll99,Floyd00} is consistent with the equation for
the conservation of probability, its trajectories are not
distributed in accordance with the density of the wave function as
in the case of the Bohm interpretation. Indeed, unlike Bohmian
trajectories, the density of Floydian trajectories would
subsequently not be that of the wave function. Rather, Floyd claims
that the necessity of the probabilistic interpretation of the wave
function in orthodox quantum mechanics arises from the incomplete
specification of a particular microstate.  Moreover, each
trajectory corresponds to an individual microstate of the
Schr\"{o}dinger equation but is nevertheless sufficient to specify
the corresponding wave function~\cite{Floyd00a}. We defer to a
later section the derivation of the kinematic equation for Floydian
trajectories and here more explicitly outline the work on the
quantum stationary Hamilton-Jacobi equation and its solutions in
one dimension.

\subsection{The quantum stationary Hamilton-Jacobi equation and its solutions}
\label{sec:qshjesol}
%[2|E]
The quantum stationary Hamilton-Jacobi equation (QSHJE) is given in
one dimension $q$ by~\cite{Floyd96,Floyd00}
\begin{equation}
\frac{(W_{\mu}')^2}{2m} + V - E = -\frac{\hbar ^2}{4m} \Dsb{W_{\mu};q}\Dsk{}
\label{eq:floyt1}
\end{equation}
where $W_{\mu}' = \frac{\partial W_{\mu}}{\partial q}$ is the
momentum conjugate to $q$ and $\Dsb{W_{\mu};q}\Dsk{} $ is the
Schwarzian derivative of $W_{\mu}$ with respect to $q$. The
Schwarzian derivative,
\begin{displaymath}
\Dsb{W_{\mu};q}\Dsk{} =
\frac{W_{\mu}'''}{W_{\mu}'}-
\frac{3}{2}\left(\frac{W_{\mu}''}{W_{\mu}'}\right)^2,
\end{displaymath}
which is unaffected by the sign of $W_{\mu}$, means that the QSHJE
is a third-order nonlinear differential equation.  The left hand
side of equation~(\ref{eq:floyt1}) replicates the classical
Hamilton-Jacobi equation, whereas the right hand side embodies the
quantum effects in the Schwarzian derivative.  Comparing
equation~(\ref{eq:floyt1}) with equation~(\ref{eq:Boht3}) one may
identify the quantum potential term as
\begin{equation}
Q= \frac{\hbar ^2}{4m}\Dsb{W_{\mu};q}\Dsk{}.
\label{eq:floyt1a}
\end{equation}
However, we recognise that the Bohm quantum potential is not always
equal to the Schwarzian derivative form.  The latter is only
guaranteed when $W_{\mu}=W$.  Equation~(\ref{eq:floyt1}) can be
obtained from the Bohm formulation for stationary systems by
recognising that, in one-dimension, equation~(\ref{eq:Boht4}) has a
solution $R= (Cm) / \sqrt{|W'|}$, where C is a constant.  (Note
that this is not generally the case in higher dimensions).
Substituting for $R$ in equation~(\ref{eq:Boht3}) one obtains the
form of equation~(\ref{eq:floyt1}).  This approach corresponds to
substituting $\psi = \exp({\pm{i W_{\mu}} / \hbar}) /
\sqrt{|W_{\mu}'|}$ (or a linear combination of the latter) into the
Schr\"{o}dinger equation.
% See Maple sheet floydvbohm for derivation
Floyd discovered that the general solution, $W_{\mu}'$, of the
QSHJE is given by
\begin{equation}
W_{\mu}'= (2m)^{1/2}(a\phi ^2+b\theta ^2+c\phi \theta )^{-1}
\label{eq:floyt2}
\end{equation}
where $[a,b,c]$ is a set of real coefficients such that $a,b
> 0$, and  $(\phi,\theta)$ must be a pair of (normalised) independent
solutions of the stationary Schr\"{o}dinger equation, $-\hbar
^2\psi''/(2m) + (V-E)\psi = 0$ for given energy $E$.  (Neither of the
the solutions $(\phi,\theta)$ is necessarily square integrable
though, as mentioned in section~\ref{sec:eqpri}, in the case of
eigenstates of bound systems one and only one is.) For given $E$,
$W_{\mu}$ is therefore prescribed by the three constants $[a,b,c]$.
Floyd has observed~\cite{Floyd94} that these constants of the
motion may alternatively be represented by the set
$[x_0,\dot{x_0},\ddot{x_0}]$ of trajectory initial conditions. This
is to be contrasted with the set $[x_0]$ required in classical
mechanics to specify $W_{\mu}$ for a given $E$ and is a reflection
of the third order nature of the QSHJE. Given the time
parameterisation provided by the kinematic equation of the Floydian
trajectories, each trajectory therefore specifies a unique
microstate $W_{\mu}$ and vice-versa.  Floydian trajectories for a
given eigenstate are therefore correspondingly more various than
Bohm trajectories. Indeed, unlike Bohm trajectories, Floydian
trajectories (for different microstates) of an eigenstates may
cross.

How do the microstates arise?  Floyd has shown~\cite{Floyd96} that
for energy eigenstates of stationary bound systems in one dimension
(for which the wave function is real), equation~(\ref{eq:floyt1})
has singular values at precisely the locations where the boundary
values are applied. This is because $W_{\mu}' \rightarrow 0$ as $x
\rightarrow \pm {\infty}$ since, whilst one of the independent
solutions of Schr\"{o}dinger equation is square-integrable, the
other must be unbounded at $x\rightarrow \pm {\infty}$ regardless
of $(a,b,c)$ given their constraints.  Thus, boundary conditions
$W_{\mu}'(x \rightarrow \infty) = 0$ admit non-unique solutions for
$W_{\mu}'$ and so yield an infinity of trajectories.  This
conclusion also applies to semi-bound systems (eg potential
barriers where the total energy is less than the barrier height).
We observe that the balance of energy between the kinetic energy
and quantum potential terms will differ between the microstates of
an eigenstate of such a bound system.  Floyd~\cite{Floyd96} also
provides evidence to suggest that an uncountable number of
microstates is also possible for bound states in dimensions greater
than one, though does not demonstrate that this is the case.

For unbound systems, in which the (complex) solution of the
Schr\"{o}dinger equation is an initial value problem, the initial
conditions determine a unique solution to
equation~(\ref{eq:floyt1}) for $W_{\mu}'$.  In this case, the same
solution is obtained in the Bohm interpretation ie $W_{\mu}=W$.
Thus, there are no microstates in this case and the trajectories,
for given initial conditions, are unique.  However, because their
kinematic equations differ, the Floydian and Bohmian trajectories
follow different paths through space-time~\cite{Floyd94}.

\subsection{General form of the wave function used for Floydian trajectories}
\label{sec:gfwf}

Given a solution $W_{\mu}$ of the QSHJE for a stationary state of
energy $E$, the corresponding wave function $\Psi$ may written in
the general bi-polar form described in section~\ref{sec:eqpri}:
\begin{equation}
\Psi = {\psi}^{+}_{\mu} + {\psi}^{-}_{\mu}
\label{eq:gfwf1}
\end{equation}
where
\begin{equation}
{\psi}^{\pm{}}_{\mu}=A^{\pm{}}_{\mu} R_{\mu}
\exp(\frac{\pm{i W_{\mu}}}{\hbar})\exp(-i\frac{Et}{\hbar}).
\label{eq:gfwf2}
\end{equation}
$A^{\pm{}}_{\mu}$ are complex constants and $R_{\mu}^2 \propto 1 /
|\nabla W_{\mu}|$ as discussed in section~\ref{sec:qshjesol}. This
form is motivated by the fact that the QSHJE is invariant under a
sign change of $W_{\mu}$, so that both ${\psi}^{\pm{}}_{\mu}$ and
any linear combination of them lead to the same QSHJE and
continuity equation when substituted into the Schr\"{o}dinger
equation.  (For a given stationary state, $\Psi$ may, of course,
also be expressed in the form of a synthesized wave~\cite{Floyd00}
using the Bohmian ansatz $\Psi = \tilde{R_{\mu}} \exp(\frac{i
\tilde{W_{\mu}}}{\hbar})\exp(-i\frac{Et}{\hbar})$.  Selection of the proper
form depends on the physical situation.)  The case $A^{-}_{\mu}=0$
can represent the complex eigenfunctions of unbound stationary
systems whereas $A^{-}_{\mu}=A^{+*}_{\mu}$ can represent the real
eigenfunctions of bound systems. Moreover, in both instances the
equivalence principle requirement that $W_{\mu}
\neq constant$ is fulfilled. Thus, the general wave function $\Psi$
comprises two waves, ${\psi}^{\pm{}}_{\mu}$, of identical form and
energy, travelling in opposite directions but scaled by different
different complex amplitudes $A^{\pm{}}_{\mu}$ respectively.  We
refer to these two waves as the positive and negative running waves
in correspondence with the sign prefix of $W_{\mu}$ in their phase
factors. An important aspect for the next section is that the
momentum and energy in the QSHJE can be derived from either of such
a pair of running waves using $p_{\mu} = \nabla S_{\mu}$ and
$E=-\partial S_{\mu} /
\partial t$ where $S_{\mu}=W_{\mu}-Et$.

\section{The total energy variational derivative of kinetic energy for stationary systems}
\label{sec:vdke}
% See [61|K]

We now investigate the variation of kinetic energy with respect to
the total energy for stationary systems. This variational
dependence is used in the definition of the kinematic equation for
Floydian trajectories and in their connection with Bohmian
trajectories. Though Floyd actually worked with the variation of
the quantum potential~\cite{Floyd82} with respect to the total
energy, it is simply related to the variation of kinetic energy
with respect total energy.  Here we find it more convenient to work
with the latter.

The use of an energy derivative implies a continuous spectrum of
admissible energy eigenvalues.  Indeed, Floyd provides examples of
Floydian trajectories (using the energy derivative of the quantum
potential) for systems with continuous energy spectra for which
such a derivative is easily defined~\cite{Floyd94}. In his work on
bound states, in which microstates of the Schr\"{o}dinger equation
are identified, Floyd~\cite{Floyd82} gives examples of microstate
trajectories in mechanical phase space for the quantum harmonic
oscillator, based on a power series solution (in the space
variable) for the modified potential (classical plus quantum
potential) which has energy depend coefficients.   For such
systems, with discrete spectra, the concept of the energy
derivative is unnatural, for the energy is not obviously a
continuous variable. Indeed, in section~\ref{sec:eqpri} we saw that
a condition for the existence of the Schwarzian derivative was that
the energy correspond to an eigenstate of the Schr\"{o}dinger
equation.  It is therefore necessary to modify the methods
in~\cite{Floyd82} to take account of the time dependence which an
energy variation between eigentstates might induce. Thus, instead
of Floyd's $\partial Q
/ \partial E$, we adopt the notation $\delta Q
/ \delta E$ described as a variational derivative with respect to
energy.

In the following sub-sections, we investigate the energy
variational derivative of the kinetic energy $\delta T / \delta E$
from the standpoint of stationary quantum systems with distinct
energy eigenstates.  The derivative (justifiably described as
partial) for continuous energy spectra is then obtained as a
limiting case of the discrete spectrum result.

\subsection{Stationary state variation}
\label{sec:pdqeb}

Let ${\Psi}_{i}$ be a stationary solution of the Schr\"{o}dinger
equation
\begin{equation}
i\hbar \frac{\partial {\Psi}_{i}}{\partial t} = H {\Psi}_{i}
\label{eq:pdqeb1}
\end{equation}
for a given potential $V(q)$ and set $\mathcal{B}$ of boundary (or
initial) conditions. So that we remain within the principles of
quantum mechanics and stay with a description of the same physical
system, we impose upon the varied wave function, $\Psi={\Psi}_{i} +
\delta \Psi$, the condition that it also satisfies
the Schr\"{o}dinger equation
\begin{equation}
i\hbar \frac{\partial \Psi}{\partial t} = H \Psi,
\label{eq:pdqeb2}
\end{equation}
for the same potential $V(q)$ and conditions $\mathcal{B}$.
%In order to conserve probability...%
We also impose the further condition that the varied wave function
preserve the space normalisation of the original wave function but
accept that the varied wave function will not necessarily be a
stationary solution of the Schr\"{o}dinger equation.

Expressing ${\Psi}_{i}$ and $\Psi$ in the bi-polar form of
section~\ref{sec:gfwf}, we obtain their respective Hamilton-Jacobi
equations whose difference may be written as
\begin{equation}
\delta \left(\frac{1}{2m} {\left(\nabla S_{\mu} \right)}^{2} \right) +
\delta Q_{\mu} = \delta \left(-\frac{\partial S_{\mu}}{\partial t}\right).
\label{eq:pdqeb3}
\end{equation}
where $S_{\mu}= \pm{W_{\mu}} - Et$. This equation may be expressed
in the simple form
\begin{equation}
{\delta T}_{\mu} + {\delta Q}_{\mu} = \delta E
\label{eq:pdqeb4}
\end{equation}
where, by identification of terms in equation~(\ref{eq:pdqeb3}),
$T_{\mu}$ is the kinetic energy and $E$ the total energy.  This is
a general equation for variations satisfying the constraints of the
Schr\"{o}dinger equation regardless of boundary conditions and
normalisation. Thus, under such general conditions, we may have
$E=E(q,t)$.  In the following, the variational derivative $\delta
T_{\mu} / \delta E$ is obtained from the ratio of the variations
${\delta T}_{\mu}$ and $\delta E$ with respect to a particular form
of infinitesimal variation of the wave function $\delta \Psi$ which
satisfies (1) the Schr\"{o}dinger equation, (2) the conditions
$\mathcal{B}$ and (3) preserves the normalisation of the wave
function.

All three conditions on the varied wave function $\Psi$ are
satisfied if it is expressed in the form
\begin{equation}
\Psi = \cos (\theta) {\Psi}_{i} + \sin (\theta) {\Psi}_{j}
\label{eq:pdqeb4a}
\end{equation}
where $\theta$ is the parameter of the variation and ${\Psi}_{j}$
is another normalised stationary solution (eigenfunction) of the
Schr\"{o}dinger equation for the same potential $V(q)$ and
conditions $\mathcal{B}$.  The orthonormality of ${\Psi}_{i}$ and
${\Psi}_{j}$ guarantees the $\theta$-independence of the
normalisation of $\Psi$.  That $\Psi$ is a linear combination of
solutions of the Schr\"{o}dinger equation for the conditions
$\mathcal{B}$ means that it also is a solution of the same.  In
principle, one can construct any solution as a linear combination
of any number of eigenfunctions. However, the objective in this
case is to study the relative variation of the kinetic and total
energies under an infinitesimal change of the wave function from
one eigenstate towards another whether or not they have a
continuous energy spectrum; hence the form of
equation~(\ref{eq:pdqeb4a}).  Whilst this analysis could be carried
out for any pair of eigenfunctions ${\Psi}_{i}$ and ${\Psi}_{j}$,
here, we define ${\Psi}_{i}$ and ${\Psi}_{j}$ to be non-degenerate
eigenstates corresponding respectively to adjacent different
eigenvalues $E_{i}$ and $E_{j}$ in the ordered energy spectrum, so
that $E_{j} > E_{i}$. In this way, we remove ambiguity from the
definition of the variational derivative with respect to total
energy and bring it into correspondence with the partial derivative
$\partial T / \partial E$ in classical conservative systems.

As discussed in section~\ref{sec:gfwf}, ${\Psi}_{i}$, ${\Psi}_{j}$
and thus $\Psi$ may be decomposed into positive and negative
running waves so that $\Psi = {\psi}^{+}_{\mu} + {\psi}^{-}_{\mu}$
with
\begin{equation}
{\psi}^{\pm{}}_{\mu} = {\psi}^{\pm{}}_{\mu i} \cos{\theta} +
{\psi}^{\pm{}}_{\mu j} \sin{\theta}
\label{eq:pdqeb4b}
\end{equation}
where
\begin{equation}
{\psi}^{\pm{}}_{\mu k} = A_{k}^{\pm{}} R_{\mu k} \exp(\frac{\pm{i
W_{\mu k}}}{\hbar}) \exp(-i\frac{E_{k}t}{\hbar})
\label{eq:pdqeb4c}
\end{equation}
are themselves the positive and negative running waves for the
microstate solution $W_{\mu k}$ of the QSHJE for the $k$th
eigenstate with eigenvalue $E_{k}$.

\subsection{Analysis of kinetic energy variational derivative}
\label{sec:avdke}

In this section, for the sake of notational clarity, we initially
drop the microstate suffix $_{\mu}$ on the variables $p$, $S$, $T$,
$\psi$ and $Q$ and re-introduce them in the final result.

Consider a variation of the wave function parameterised by $\delta
\theta$. From equations~(\ref{eq:pdqeb3}) and~(\ref{eq:pdqeb4}), the
change in the kinetic energy is
\begin{equation}
{\delta T} = \frac{p}{m}\frac{\partial p}{\partial
\theta}
\delta \theta
\label{eq:avdke1}
\end{equation}
where $p = \nabla S$  is the conjugate momentum. Similarly, the
change in the total energy is
\begin{equation}
\delta E = \frac{\partial E}{\partial \theta} \delta \theta
\label{eq:avdke2}
\end{equation}
where $E = - \partial S / \partial t$ is the total energy. Thus, in
this work, the variational derivative of the kinetic energy with
respect to the total energy is defined by
\begin{equation}
\frac{\delta T}{\delta E} = \frac{1}{m}
\frac{p \cdot \partial p / \partial \theta}{\partial E / \partial \theta}.
\label{eq:avdke3}
\end{equation}
In both the latter equations the partial derivatives with respect
to $\theta$ are evaluated at $\theta = 0$ ie in the unvaried state.
In passing, we note from equation~(\ref{eq:pdqeb4}) that ${\delta
Q} / \delta E = 1 - \delta T / \delta E$.

The subsequent analysis proceeds by respectively expressing the
conjugate momentum and total energy in terms of one of the running
waves (we choose the positive wave) of the varied wave function
(see equation~(\ref{eq:pdqeb4b})), so that
\begin{eqnarray}
&&p = \nabla S = \Im \left ( \frac{\hbar}{{\psi}}
\nabla
{\psi}
\right) \cr
&&E = -\frac{\partial S}{\partial t} = - \Im \left (
\frac{\hbar}{{\psi}} \frac{\partial {\psi}}{ \partial t}
\right)
\label{eq:avdke4}
\end{eqnarray}
in which, and hereafter, the $\pm{}$ superscript has been dropped.
Therefore,
\begin{eqnarray}
&& \frac{\partial p}{\partial \theta} =
\hbar \Im \left[
-\frac{1}{{\psi}}\frac{\partial {\psi}}{\partial \theta}
\frac{1}{{\psi}}\nabla {\psi}
+ \frac{1}{{\psi}}\frac{\partial}{\partial \theta} \nabla {\psi}
\right] \cr
&& \frac{\partial E}{\partial \theta} =
-\hbar \Im \left[
-\frac{1}{{\psi}} \frac{\partial {\psi}}{\partial \theta}
\frac{1}{{\psi}} \frac{\partial {\psi}}{\partial t}
+ \frac{1}{{\psi}}\frac{\partial}{\partial \theta}
\frac{\partial {\psi}}{\partial t}
\right]
\label{eq:avdke5}
\end{eqnarray}
where the partial derivative terms with respect to $\theta$ at
$\theta = 0$ are computed from equation~(\ref{eq:pdqeb4b}) (without
the superscripts) as
\begin{eqnarray}
&& \left( \frac{1}{{\psi}} \frac{\partial {\psi}}{\partial
\theta}
\right )_{\theta = 0} =
\frac{{\psi}_{ j}}{{\psi}_{ i}} \cr
&& \left ( \frac{1}{{\psi}} \frac{\partial}{\partial \theta}
\nabla {\psi} \right )_{\theta = 0} =
\frac{1}{{\psi}_{ i}} \nabla {\psi}_{ j} \cr
&& \left ( \frac{1}{{\psi}} \frac{\partial}{\partial
\theta}
\frac{\partial {\psi}}{\partial t} \right )_{\theta = 0} =
\frac{1}{{\psi}_{ i}} \frac {\partial {\psi}_{ j}}{\partial t}.
\label{eq:avdke6}
\end{eqnarray}
With these results equations~(\ref{eq:avdke5}) become
\begin{eqnarray}
&& \left( \frac{\partial p }{\partial \theta} \right)_{ij} =
\hbar \Im \left[
-\frac{{\psi}_{ j}}{{\psi}_{ i}}
\left(
\frac{\nabla {\psi}_{ j}}{{\psi}_{ j}}
- \frac{\nabla {\psi}_{ i}}{{\psi}_{ i}}
\right)
\right] \cr
&& \left( \frac{\partial E}{\partial \theta} \right)_{ij} =
- \hbar \Im \left[
-\frac{{\psi}_{ j}}{{\psi}_{ i}}
\left(
\frac{1}{{\psi}_{ j}} \frac{\partial {\psi}_{ j}}{\partial t}
- \frac{1}{{\psi}_{ i}} \frac{\partial {\psi}_{ i}}{\partial t}
\right)
\right],
\label{eq:avdke7}
\end{eqnarray}
in which the partial derivatives with respect to $\theta$ at
$\theta =0$ have been labelled by the suffices $ij$ to show that
they are evaluated at the eigenstate $i$ `in the direction of' the
eigenstate $j$.  Expressing the running wave for the $k$th
eigenstate in the form of equation~(\ref{eq:pdqeb4c}) and
re-introducing the microstate suffix $\mu$,
equations~(\ref{eq:avdke6}) may be used in
equation~(\ref{eq:avdke3}) to finally give
\begin{eqnarray}
&&\left( \frac{\delta T}{\delta E} \right )_{\mu ij} =
\frac{\nabla S_{\mu i}}{m \Delta E_{ji}} \cdot \nonumber \\
&&
\left[
\nabla (\Delta S_{\mu ji})
 + \hbar \tan \left( \frac{\Delta S_{\mu ji}}{\hbar} + \Delta
{\alpha}_{\mu ji} \right)
\nabla \ln \left( \frac{R_{\mu j}}{R_{\mu i}} \right)
\right],
\label{eq:avdke8}
\end{eqnarray}
where $\Delta E_{ji} = E_{j} - E_{i}$, $\Delta S_{\mu ji} = S_{\mu
j} - S_{\mu i}$ and $S_{\mu k}(q,t)=W_{\mu k}(q)-E_{k}t$.  $\Delta
{\alpha}_{\mu ji} = {\alpha}_{\mu j} - {\alpha}_{\mu i}$ is the
difference between the phases of the running wave scale factors of
the $j$th and $i$th eigentsates for microstate $_{\mu}$, where
$A_{\mu k}=|A_{\mu k}| \exp{(i {\alpha}_{\mu k})}$.

We recall that this analysis was done for the positive running wave
but observe that the result~(\ref{eq:avdke8}) is invariant to a
change of sign prefix to $W_{\mu}$ (for the negative running wave)
as expected if the scale factor phases also undergo a corresponding
sign change. The latter condition does indeed occur when the
eigenstates are real as in the case of bound states. (See
section~\ref{sec:gfwf}.) Equation~(\ref{eq:avdke8}) clearly shows
that $\left( {\delta T}/{\delta E} \right )_{\mu ij}$ may be
explicitly time dependent on account of the 'beating' of the
different frequencies of the states $i$ and $j$.

\subsection{The kinetic energy variational derivative for unbound systems}
\label{sec:kvdubs}

In the limit of unbound systems (including barriers), which have
continuous energy spectra, $\Delta S_{\mu ji}$ and $\Delta E_{ji}$
become infinitesimals and $R_{\mu j} \rightarrow R_{\mu i} $ so
equation~(\ref{eq:avdke8}) becomes
\begin{equation}
\left( \frac{\delta T}{\delta E} \right )_{\mu ij} \rightarrow
\left( \frac{\delta T}{\delta E} \right )_{\mu} =
\frac{\nabla {S}_{\mu}}{m} \cdot
\nabla \left( \frac{\partial S_{\mu}}{\partial E} \right),
\label{eq:avdke9}
\end{equation}
where the right hand side is identical to the classical form.

In the case of a free system in a constant potential, the spectrum
is continuous and $R$ is a constant so that the quantum potential
is zero. We may then replace $\delta T / \delta E$ by $\partial T /
\partial E$ in equation~(\ref{eq:avdke9}) and obtain precisely the
classical result.  Clearly, for unbound systems, the energy
derivative of the kinetic energy has no explicit time dependence.

\subsection{The kinetic energy variational derivative for bound systems}
\label{sec:kvdbs}

In the case of stationary bound systems with discrete spectra and
real eigenfunctions, equation~(\ref{eq:avdke8}) may be applied to
any one of the corresponding Floydian microstates of the
Schr\"{o}dinger equation. Referring to equation~(\ref{eq:floyt2})
and the text immediately thereafter, a microstate may be identified
by the set $[q_0,\dot{q_0},\ddot{q_0}, E_{i}]$ of trajectory
initial conditions and the energy eigenvalue. Thus, to determine
from equation~(\ref{eq:avdke8}) the kinetic energy variational
derivative for the microstate of the $i$th eigenstate defined by
the initial conditions $[q_0,\dot{q_0},\ddot{q_0}]$, we may define
$\Delta W_{\mu ji}(q)=W_{\mu}(q,[q_0,\dot{q_0},\ddot{q_0},
E_{j}])-W_{\mu}(q,[q_0,\dot{q_0},\ddot{q_0}, E_{i}])$ holding the
trajectory initial conditions constant.  For such microstates, the
conjugate momentum $\nabla S_{\mu i}$ is non-zero and the second
term within the square bracket of equation~(\ref{eq:avdke8})
generates the explicit time dependence of $\left( {\delta
T}/{\delta E}\right )_{\mu ij}$ with a frequency equal to the
difference between the frequencies of the states $i$ and $j$.  We
therefore observe that for a given set of initial conditions there
is a family of explicitly time dependent trajectories corresponding
to the family of discrete eigenstates.  The explicit time
dependence of such bound state microstate trajectories does not
appear in the work of Floyd.

\section{A trajectory equation from the energy variational derivative of the action}
\label{sec:fleq}

\subsection{The nature of the energy variational derivative}
\label{sec:fleqn}

Here we study the basis of the Floydian trajectory method as
defined for stationary systems. The approach, in the augmented
domain of the motion $(q,t,E)$, is based upon the variation of the
action with respect to changes in the total energy $E$ which are
out of the constant energy plane of the motion. As such, it is
restricted to stationary systems. We emphasise that the kinematic
equations so obtained are for Floydian trajectories in the $(q,t)$
domain.

Here we make a point of clarification.  Carroll~\cite{Carroll99}
connects the evaluation of the total energy derivative of the
action with the Legendre transformation from the $(q,t)$ domain to
the $(q,E)$ domain. In the case of stationary systems, such a
transformation corresponds to the substitution of one independent
variable, $t$, by another, $E$, which is a single valued constant
of the motion. Thus, the trajectory in $(q,E)$ space would
trivially lie on a fixed energy plane labelled by $E$. (In the case
of non-stationary systems the Legendre transformation is well
defined and the motion is not restricted to such energy planes.) We
emphasise that the Legendre transformation of a given motion
between the $(q,t)$ and $(q,E)$ domains is to be clearly
distinguished from the following study of variations of the motion
in the $(q,t,E)$ domain.  We now proceed with the latter.  We once
again drop the microstate suffix $\mu$ and the eigenstate suffices
$ij$, using them only when clarity demands.

\subsection{The energy variational derivative of the action and a trajectory kinematic equation}
\label{sec:edrac}

Consider the energy variational derivative of the (quantum or
classical) action $S(q,t,E)$ which, following Floyd, we define as
the epoch
\begin{equation}
\tau = -\frac{\delta S}{\delta E}.
\label{eq:fleq1}
\end{equation}
In using the variational derivative $\delta S
/ \delta E$ we are consistent with section~\ref{sec:vdke} but
differ from Floyd, who uses the partial derivative.  The use of the
latter is justified only in the case of stationary classical and
quantum systems which have a continuum of states of spatially
uniform energy.  Our use of the variational derivative acknowledges
that in quantum mechanics the variation $\delta E$ may necessarily
be a function of $q$ and $t$ in order for the Schr\"{o}dinger
equation and its boundary or initial conditions to remain
satisfied.

We suspend Floyd's assumption that $\tau$ is a constant and examine
its time derivative along a trajectory in $(q,t)$ space in the
energy plane labelled by $E$. Thus, from equation~(\ref{eq:fleq1}),
\begin{equation}
\frac{ d\tau}{d t} = - \frac{d}{dt}\left(\frac{\delta S}{\delta E} \right)
= -\left[
\frac{\partial}{\partial t}\left(\frac{ {\delta} S}{\delta
E}\right) +
\nabla \left( \frac{\delta S}{\delta E}\right) \cdot \dot{q}
\right].
\label{eq:fleq2}
\end{equation}
Changing the order of $\partial t$ and $\delta E$ and using the
Hamilton-Jacobi equation with $E = - \partial S /
\partial t$ (see equation~(\ref{eq:pdqeb4})),
equation~(\ref{eq:fleq2}) becomes
\begin{equation}
\nabla \left( \frac{\delta S}{\delta E} \right) \cdot
\frac{\dot{q}}{(1 - d \tau / d t)} = 1
\label{eq:fleq4}
\end{equation}
where $\dot{q}$ is the total derivative of the trajectory position
with respect to time.

A further equation involving $\delta S
/ \delta E$ is obtained from the energy variational
derivative of the kinetic energy $T = (\nabla S)^{2}
/ 2m$.  Thus, after changing the order of $\delta$
and $\nabla$ as applied to $S$,
\begin{equation}
\nabla \left(\frac{\delta S}{\delta E}\right) \cdot
\frac{\nabla S}{m (\delta T / \delta E)} = 1,
\label{eq:fleq5}
\end{equation}
where the $ij$ subscripts, dropped from $(\delta T / \delta
E)_{ij}$, are understood.  Comparison of equations~(\ref{eq:fleq4})
and~(\ref{eq:fleq5}) shows that equation~(\ref{eq:fleq4}) is
satisfied by
\begin{equation}
\dot{q} = \frac{\nabla S }{m}\frac{(1 - d\tau / dt)}{(\delta T / \delta E)}.
\label{eq:fleq6}
\end{equation}
Whilst this solution is consistent with the one-dimensional case,
for which it is the only solution, in higher dimensions it is not
the only solution of equation~(\ref{eq:fleq4}). For an
$n$-dimensional system, equation~(\ref{eq:fleq4}) defines $\dot{q}
/ (1 - d\tau / d t)$ as a member of an infinite set, ${\mathcal{Q}}_{n-1}$, of vectors
passing from the origin and ending on an $(n-1)$-dimensional plane
which has a normal vector $\nabla ({\delta S} / {\delta E})$ and
which is located at a distance $1 / (|{\nabla \left({\delta S} /
{\delta E}\right)}|)$ from the origin. For such a system,
equation~(\ref{eq:fleq5}) shows that ${\nabla S}
/ {(m ({\delta T}/{\delta E}))} \in
{\mathcal{Q}}_{n-1}$ also. By choosing $\dot{q} / (1 - d\tau / d
t)$ to be this particular member of ${\mathcal{Q}}_{n-1}$, we
obtain equation~(\ref{eq:fleq6}) and achieve consistency with the
1-dimensional case. Thus, in all except the 1-dimensional case, the
form of the trajectory equation~(\ref{eq:fleq6}) is suggested,
rather than prescribed, by the above variational analysis.

We note that in this derivation, no use has been made of the
probability conservation equation~(\ref{eq:Boht4}). Thus, unlike
Bohmian trajectories, the trajectories defined by
equation~(\ref{eq:fleq6}) do not have a natural probabilistic
interpretation.  We discuss this matter further in the next
section.

\section{Connections between classical, Floydian and Bohmian trajectories}
\label{sec:conbf}
% [56 | K]

The Floydian and Bohmian trajectories are connected by different
assumptions about the time dependence of the parameter $\tau$. In
classical mechanics, this parameter is demonstrably a constant with
respect to the time $t$.  This is not the case in quantum
mechanics.

\subsection{Classical systems}
\label{sec:clasys}

For conservative classical systems, $\delta T / \delta E=1$ and in
cartesian coordinates $m\dot{q}=\nabla S$ so
equation~(\ref{eq:fleq6}) implies that ${d\tau}
/ {dt} = 0$ ie $\tau$ is indeed a constant without assumption.  On
the other hand, the converse is not true: Assuming that ${d\tau} /
{dt} = 0$ is a necessary though not sufficient condition for
$m\dot{q}=\nabla S$ except in the case of one space dimension, as
discussed in section~\ref{sec:edrac}.

\subsection{The kinematic equation of Floydian trajectories}
\label{sec:kmflt}

The kinematic equation of Floydian trajectories arises from the
{\it assumption} that ${d\tau} / {dt} = 0$.   Thus, with this
assumption equation~(\ref{eq:fleq6}) becomes
\begin{equation}
\dot{q} = \frac{\nabla S }{m}\frac{1}{(\delta T / \delta E)},
\label{eq:kmflt1}
\end{equation}
in which we may use the expression for $(\delta T / \delta E)_{ij}$
given in equation~(\ref{eq:avdke8}).

If we explicitly write the action in the stationary form $S = W -
Et$ so that $\nabla S = \nabla W$ and recall from
equation~(\ref{eq:pdqeb4}) that $\delta T / \delta E = 1 - \delta Q
/ \delta E$, then equation~(\ref{eq:kmflt1}) becomes
\begin{equation}
\dot{q} = \frac{\nabla W }{m}\frac{1}{(1 - \delta Q / \delta E)},
\label{eq:kmflt2}
\end{equation}
which is precisely the form used by Floyd~\cite{Floyd82}.  For
(Floydian) microstate trajectories defined by this kinematic
equation, $\dot{q} \neq \frac{p}{m}$ where $p =  \nabla W$ is the
conjugate momentum. This is different from the kinematic relation
for both classical and Bohm trajectories, the latter applying
directly to the reduced action $W$ of the wave function and not
microstate solutions $W_{\mu}$.

In passing, we note that equation~(\ref{eq:kmflt2}) represents a
deformation of the classical kinematic relation between conjugate
momentum and velocity.  In making a parallel with relativity,
Faraggi and Matone~\cite{Matone00a} capture this deformation in the
form of a 'quantum mass field'
\begin{displaymath}
m_Q = m (1 - \delta Q / \delta E) = m \delta T / \delta E,
\end{displaymath}
so restoring the classical form $\dot{q} = p/m_Q$.  Only for
microstate solutions of bound states, for which $\delta T / \delta
E$ is not uniformly zero, is the 'quantum field mass' a meaningful
concept.  The multiplicity of such microstates implies an equal
multiplicity of 'quantum mass fields'.

\subsection{The kinematic equation of Bohmian trajectories
and quantum time}
\label{sec:kmbmt}

The kinematic equation of Bohmian trajectories is derived from the
conservation of probability equation and the classical form for the
current density.  This unambiguously implies that $ \dot{q} =
\frac{1}{m} \nabla S$ in the coordinate representation so coinciding with the
classical kinematic relation.  (A different kinematic equation
occurs in other representations of the Bohm
formulation~\cite{Brown00}.) The classical and Bohm kinematic
relations therefore coincide. Thus, rather than being derived from
an assumption regarding $\tau$, the Bohm kinematic equation may be
used to define the derivative $d\tau
/ dt$.  In the the coordinate representation,
equation~(\ref{eq:fleq6}) implies that
\begin{equation}
\frac{d \tau}{dt} = \frac{\delta Q}{\delta E}
\label{eq:confb8}
\end{equation}
showing that, in general, the 'epoch' $\tau$ is not a constant for
a Bohmian trajectory.  Defining the 'quantum time' as $t_{Q} = t -
\tau$, we find that
\begin{equation}
\frac{d t_{Q}}{dt} = (1 - {{\delta Q} / {\delta E}}) = {\delta T} / {\delta E},
\label{eq:confb9}
\end{equation}
along the trajectory. This equation thus suggests that the quantum
potential acts to deform time. The kinematic equation for Bohm
trajectories with respect to $t_{Q}$ is then
\begin{equation}
\frac{dq}{d t_{Q}} =
\frac{\nabla S}{m({{\delta T} / {\delta E}})}
\label{eq:confb10}
\end{equation}
which, for stationary systems, has the same form as
equation~(\ref{eq:kmflt1}) for Floydian trajectories with respect
to the classical $t$.

We observe from equation~(\ref{eq:confb9}) that for conservative
classical systems, $t_{Q}$ and $t$ advance at the same rate and, at
most, differ by an arbitrary constant shift $\tau$. This is
consistent with section~\ref{sec:clasys} and the idea that a
classical particle has neither internal energy nor extension in
space-time.

In the case of unbound quantum systems~\cite{Floyd94}, there is a
single microstate which can be compared with a Bohmian trajectory.
In this instance, the above equation~(\ref{eq:confb10}) suggests
that, in a region of interference where the quantum potential is
non-zero, the differing forms of the Floydian and Bohmian
trajectories can be related by a deformation of time which is
determined by the quantum potential.

For stationary quantum systems the square integrable real wave
functions (such as in bound states), have $\nabla W
= \nabla S = 0$ and ${{\delta Q} / {\delta E}}= 1$, so neither Bohm
trajectories nor quantum time $t_{Q}$ evolve with respect to the
classical time $t$; they stand still.  In fact for such systems,
equation~(\ref{eq:confb10}) is not well defined, this being a
symptom the fact that the basis of Bohm trajectories is
inconsistent with that of the microstates of bound state wave
functions.

\subsection{Applications to different potentials}
\label{sec:appot}

To illuminate the general results of the previous sections, we
briefly discuss their application to different forms of potential.

In the case of unbound potentials, results are obtained for both
Floydian and Bohmian trajectories and the Floydian $W_{\mu}$ and
the Bohmian $W$ are the same. For constant potentials classical
results are obtained.  For the step-potential which is less than
the total energy $E$~\cite{Floyd94}, there are points in the
interference domain where the energy derivative of the kinetic
energy passes through zero and changes sign so yielding Floydian
trajectories that have infinite positive and negative velocities.
% See MAPLE 6 work sheet floydcheck.mws
This is in distinction to the positive Bohmian velocities in the
same domain.

In the case of bound potentials~\cite{Floyd82} Floydian and Bohmian
trajectories differ not only because of differences in the
kinematic equations of their trajectories but also because
$W_{\mu}$ and $W$, from which the are respectively derived, differ
both in form and multiplicity.  For bound potentials, Floydian
trajectories are found, in this work, to have explicit time
dependence which is at the beat frequency between adjacent
eigenstates whereas Bohmian trajectories have zero velocity. Though
not explicitly addressed here, we would also expect this principle
of explicit time dependence to apply to finite bound potentials
such as the finite square well~\cite{Floyd00a} where $E < V_{0}$
(the well depth).

\section{Conclusions}
\label{sec:sumbf}

The emphasis of this paper has been technical rather than
philosophical and interpretational.  Having briefly reviewed the
Floydian trajectory method in quantum mechanics, we focused
attention upon basis of the kinematic equation of Floydian
trajectories.

Firstly, motivated by concerns over the definition of the total
energy derivative of the kinetic energy and thus the quantum
potential, especially in the case of the quantum systems with
discrete energy spectra, we derived an explicitly time dependent
form for the energy derivative of the kinetic energy for use in the
kinematic equation of Floydian trajectories.  This explicit
periodic time dependence, common to all microstates of a given
eigenstate and a modification of the kinematic equation used by
Floyd, is at the beat frequency between two adjacent eigenstates of
a bound system and is therefore itself quantised. In the limiting
case of systems with continuous spectra, this modified kinematic
equation becomes the original Floydian form.  Our approach is based
on the use a total energy variational derivative of the kinetic
energy derived from a variation of the wave function constrained to
be in the space spanned by the eigenfunctions of the
Schr\"{o}dinger equation.  This is in contrast to the approach of
Floyd who, treating the total energy as a parameter of each
microstate, directly computes the alternative total energy partial
derivative of the quantum potential without consideration of the
temporal interference with other eigenstates induced by a variation
in energy.

Secondly, we have examined the derivation of the kinematic equation
of Floydian trajectories from the energy derivative of the action
(the epoch) by first setting aside Floyd's assumption that the
epoch is a constant and working without the restriction to one
dimension.  We find that in general, whilst the magnitude of the
velocity of the Floydian trajectory is well defined, its direction
is not; the exception being in one-dimension or in systems with
uncoupled degrees of freedom.  This arbitrariness of direction is
overcome by assuming the multi-dimensional form of the kinematic
equation to be consistent with that in one dimension.  We show that
in classical mechanics the epoch is a constant without assumption,
whereas assuming it is so in quantum mechanics leads the kinematic
equation of Floydian trajectories.  On the other hand, assuming the
Bohm kinematic equation we find the epoch not to be constant and
are led to the idea of the deformation of time by the quantum
potential.  In the case of unbound systems, for which only is this
a consistent interpretation, we find that Floydian and Bohmian
trajectories can be related by a time deformation.

Further work is required to investigate the application of the
explicitly time dependent form of the total energy derivative of
the kinetic energy to the various bounded potentials already
analysed by Floyd.

\section*{Acknowledgement}
\label{sec:ackn}

The author gratefully acknowledges the helpful comments provided by
M. Matone and by E. R. Floyd in response to an earlier version of
this paper. Some points raised by E. R. Floyd are the subject of
further investigation.

% now the references. delete or change fake bibitem. delete next three
% lines and directly read in your .bbl file if you use bibtex.
% ------------------------------------------------------------------------

% ------------------------------------------------------------------------
\end{document}